\documentstyle[12pt,aaspp4]{article}

\begin{document}

\def\thefootnote{\fnsymbol{footnote}}

\title{Difference Image Analysis of the OGLE-II Bulge Data\footnotemark . I.\\
The Method.}

\author{P. R. Wo\'zniak}

\affil{Princeton University Observatory, Princeton, NJ 08544--1001, USA}
\affil{wozniak@astro.princeton.edu}

\begin{abstract}

I present an implementation of the difference image photometry based on the
Alard \& Lupton optimal PSF matching algorithm. The most important feature
distinguishing this method from the ones using Fourier divisions is that
equations are solved in real space and the knowledge of each PSF is not required
for determination of the convolution kernel. I evaluate the method and software
on 380 GB of OGLE-II bulge microlensing data obtained in 1997--1999 observing
seasons. The error distribution is Gaussian to better than 99\% with the
amplitude only 17\% above the photon noise limit for faint stars. Over the entire
range of the observed magnitudes the resulting scatter is improved by a factor
of 2--3 compared to DoPhot photometry, currently a standard tool in
microlensing searches. For testing purposes the photometry of $\sim4600$
candidate variable stars and sample difference image data are provided for
BUL\_SC1 field. In the candidate selection process, very few assumptions have
been made about the specific types of flux variations, which makes this data
set well suited for general variability studies, including the development
of the classification schemes.

\end{abstract}

\keywords{techniques: photometric --- methods: data analysis}

\footnotetext{Based on observations obtained with the 1.3 m Warsaw
Telescope at the Las Campanas Observatory of the Carnegie
Institution of Washington.}

\section{Introduction}

Microlensing in the Galaxy is an intrinsically rare phenomenon.
It happens to a couple of stars per million at any given time and this
is why Galactic microlensing surveys are monitoring millions of stars
in the densest fields of the sky: Galactic Center region and galaxies of the
Local Group which are, at least partially, resolved into stars. Paczy\'nski (1996)
presents a review of the basic theory and microlensing surveys.

The price for reasonably high event rates is complicated systematics and
limitations of the photometry in crowded fields. Overlapping stellar images
make it hard to estimate the point spread function (PSF) and inevitably influence
light centroid of the variables and number of detected sources. Over past few
years it has become clear that the optical depth to microlensing cannot be
reliably determined until the effects of blending are considered
(Nemiroff 1994, Han 1997, Wo\'zniak \& Paczy\'nski 1997). From the very
beginning microlensing surveys are upgrading their photometry and detection
techniques. In this area image subtraction is the most promising method, as it
naturally removes numerous problems by eliminating multi-PSF fits. It is often
referred to as the Difference Image Analysis (DIA) and we should probably
settle on this terminology.

There are a number of implementations based on the standard PSF matching
algorithms which involve Fourier division (Crotts 1992, Phillips \& Davis 1995,
Tomaney \& Crotts 1996, Reiss et al. 1998, Alcock et al. 1999a). These
techniques have become fairly sophisticated. Recently a nearly optimal
algorithm have been found (in a least square sense). Alard \& Lupton (1998)
eliminated division in Fourier space and came up with the technique
which is particularly well suited for crowded fields. It actually works
better in denser frames up to an enormous level of crowding, at which the
light distribution becomes almost smooth. Alard (2000) generalized this result
for spatially variable kernels.

Using a modified version of DoPhot (Schechter, Mateo \& Saha 1995,
Udalski, Kubiak \& Szyma\'nski 1997) OGLE-II has successfully detected events
in real time as well as from general searches of the database. During 3
observing seasons between 1997 and 1999 OGLE detected 214 microlensing events
(Udalski et al. 2000). However for the derivation of the optical depth it is
essential to have as much control over systematics as possible. A complete
reanalysis of the OGLE-II bulge data using image subtraction is under way.
This paper is a technical description of the implementation of software used
to perform our photometry on difference images. The catalog of $\sim500$ microlensing events
and statistical analysis will be published elsewhere.

In the remainder of this paper we describe all the steps of the data reduction
process, the software, some basic evaluations of its performance, and the
availability of the data.

\section{Overview of the photometric method}

Retrieving photometric information from images of crowded stellar
fields is an important but at the same time a difficult task. The most
serious complications are associated with overlapping stellar images.
In such conditions it is virtually impossible to get a reliable background
estimate, PSFs are ill defined, there are degeneracies in multi-parameter fits,
and finally the centroids of the light for variable stars are influenced
by neighboring stars. Any attempt to cross identify faint sources is bound
to lead to a high confusion rate. For years, observers handled this problem
using DoPhot software (Schechter, Mateo \& Saha 1995), usually customized for
a particular experiment. That package employs the traditional approach, that is
the modeling of the heavily blended neighborhood for each star, and indeed,
stands behind most of the important scientific results from microlensing so far.
Various authors have attempted subtracting images of stellar fields over the past
decade to eliminate fitting of multi-PSF models, however successful applications
were usually limited to best quality data sets and focussed on one particular type
of project. The demands encountered in microlensing surveys triggered new
efforts in this area. Several groups are now using image subtraction algorithms
based on convolution kernels derived from high signal to noise PSFs.
The basic equation for this method would be:

\begin{equation}
\hbox{\it Ker} =
\hbox{\it FFT}^{-1}
\left({\hbox{\it FFT}(PSF_1)}\over{\hbox{\it FFT}(PSF_2)}\right).
\end{equation}

A variation of this algorithm uses the above equation for the core of the PSF
and supplements this with the analytic fit in the wings, where otherwise noise
dominates the solution (e.g. Tomaney \& Crotts 1996). This technique
produced a number of results, but we believe it still suffers from some of the
problems mentioned above. The derived kernel is obviously as weak as both PSFs;
Fourier division is uncertain and difficult to control; and the more crowding
the worse it gets.

Recently an algorithm has been proposed in which the final difference
of two images of the same stellar field is nearly optimal (Alard \& Lupton 1998).
The basic idea is to work on full pixel distributions of both images
and do the calculation in real space:

\begin{equation}
\hbox{\it Im\/}(x,y) = \hbox{\it Ker\/}(x,y;u,v) \otimes \hbox{\it Ref\/}(u,v) +
\hbox{\it Bkg\/}(x,y),
\end{equation}

where $\hbox{\it Ref}$ is a reference image, $\hbox{\it Ker}$ is a convolution kernel,
$\hbox{\it Bkg}$ is a difference in background and $\hbox{\it Im}$ is a program image.
The above equation should be understood in the least square sense and treats
PSF gradients. To solve for the PSF matching kernel and background we minimize
the squared differences between the images on both sides of the Equation~2, summed
over all pixels. It is assumed that most stars do not vary, and as a result,
most pixels vary only slightly due to seeing variations. The problem is linear
for kernels made of Gaussians with constant sigmas and modified by polynomials.
For the full description of the algorithm see Alard (2000). Here we would
like to emphasize that the knowledge of the PSF and background for individual
images is not required and the method works better as the crowding increases,
because in denser fields more pixels contain information about the PSF difference.
It is very easy to impose flux conservation and the flux scale is
automatically adjusted, so that the effects of variable atmospheric extinction
and exposure time are taken out. Also, after correct subtraction the derived
centroid of the variable object is unbiased by surrounding objects, as the
variable part of the image is uncrowded. On the down side, the variables must
be found before the actual measurement and the method requires some preliminary
processing. Pixel grids of all images must be matched and images must be
resampled. Preparation of the reference image to be subtracted from all the
other frames takes some effort, and is an absolutely critical factor for the
quality of the final result.

The DIA method measures flux differences between the frames, also called the AC
signal, as opposed to the DC signal, that is the total flux, given by most
photometric tools. Intuitive arguments that measuring AC signal is inferior to
having DC signal are a common misconception, at least in microlensing. It is
certainly true that for some applications we need to know the total flux,
not just the variable part of it. However, if we can be sure of
our identification of the variable with an object seen on the reference frame,
than we can calibrate the light curve on DC scale and the result will not be
worse than using say DoPhot on the reference image. It is often merely
an illusion that we know what has varied in fields as crowded as Galactic
bar or globular clusters, and jumping to the conclusion that the DoPhot
light curve represents the truth, the whole truth and nothing but the truth,
is not advisable. At a crowding level of one source per couple of beams, source
confusion is as common as correct identification (Hogg 2000). This is the
essence of the blending problem in microlensing searches.

\subsection{The data}

A few words on the data are due in order to put the discussion of the processing and
photometric accuracy in context. All frames used in this paper were obtained
with the 1.3 m Warsaw Telescope at the Las Campanas Observatory, Chile, which
is operated by the Carnegie Institution of Washington. The ``first generation''
camera uses a SITe~3~~ $2048 \times 2048$ CCD detector with $24 \mu m$ pixels
resulting in 0.417 arcsec/pixel scale. Images of the Galactic bulge are taken in
driftscan mode at ``medium'' readout speed with the gain 7.1 $e^-$/ADU and
readout noise of 6.3 $e^-$. Saturation level is about 55,000 ADU. For the
details of the instrumental setup, we refer the reader to
Udalski, Kubiak \& Szyma\'nski (1997).

The majority of frames was taken in the $I$ photometric band. During observing
seasons of 1997--1999 the OGLE experiment collected typically between 200 and 300
$I$-band frames for each of the 49 bulge fields SC1--SC49 (for simplicity the
prefix BUL\_ will be omitted in field designations). The number of frames in $V$
and $B$ bands is small and we do not analyze them with the DIA method. The
median seeing is 1.3 arcsec for our data set.

\section{Photometric pipeline}

We start with a general description of the data flow followed by more
detailed descriptions of individual image processing algorithms in Sections~3.1
through 3.10.
For better orientation we provide schematic diagrams of the data flow in
Figures~1 and 2. Although I wrote all of the software from scratch I was strongly
inspired by programs from Alard (2000) distributed on the web at
{\it http://www.iap.fr/users/alard/package.html}.

\begin{figure}
\plotfiddle{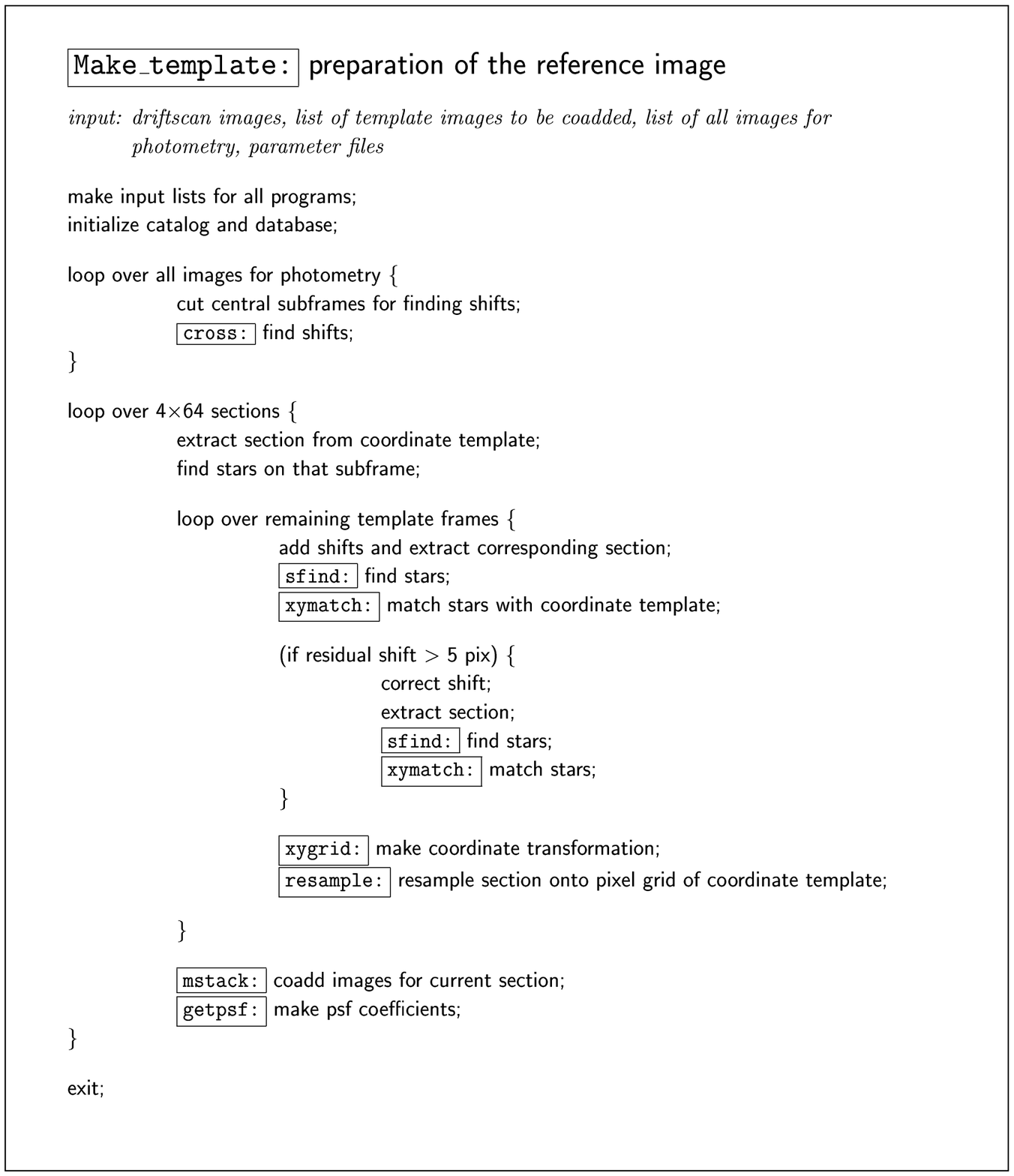}{20cm}{0.}{100.}{100.}{-320}{-100}
\caption{Construction of the reference image. Pseudo coding of the data flow.
Framed names indicate programs described in the following sections.
}
\end{figure}

\begin{figure}
\plotfiddle{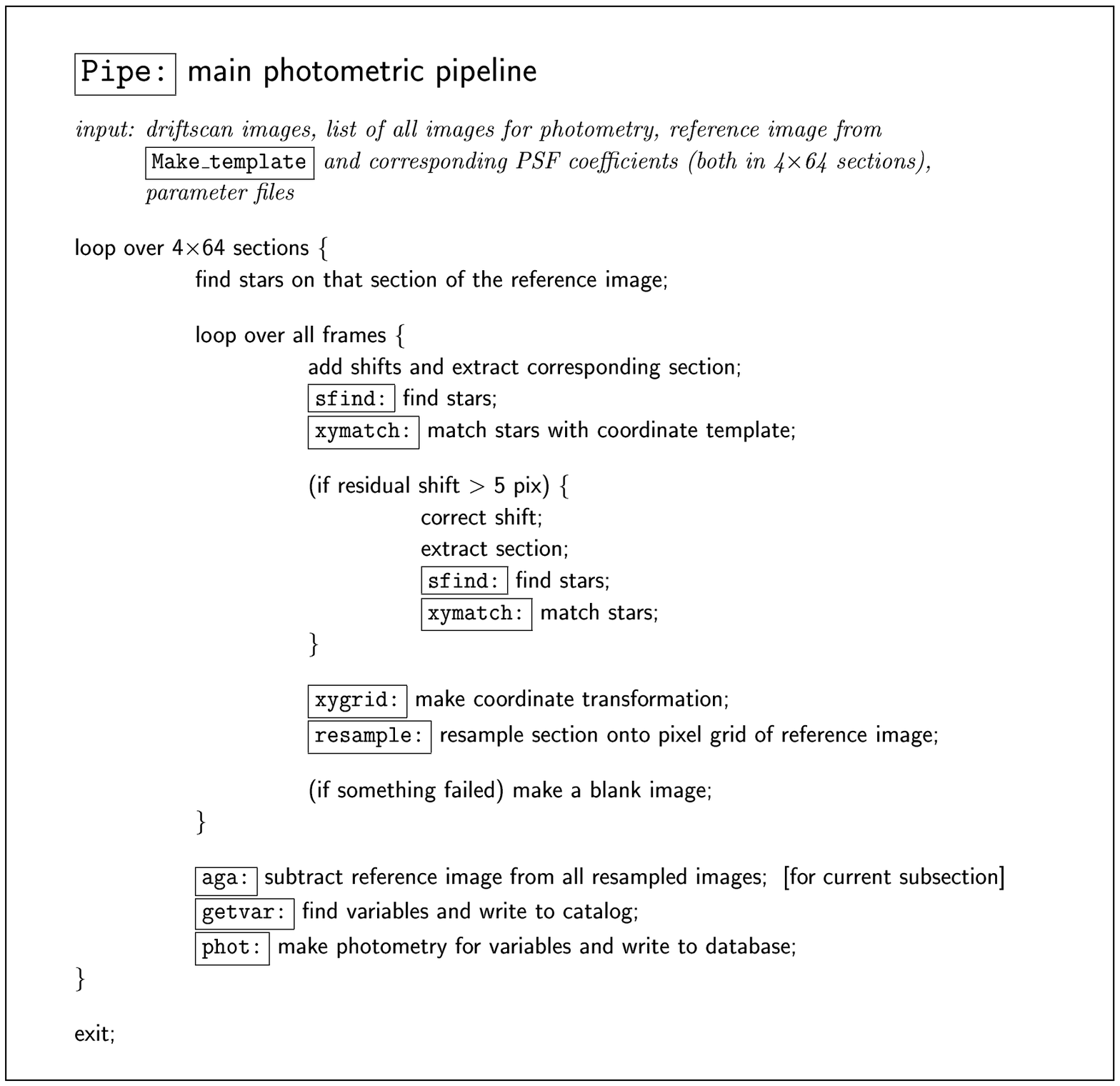}{16cm}{0.}{100.}{100.}{-320}{-170}
\caption{Main photometric pipeline. Pseudo coding of the data flow.
Framed names indicate programs described in the following sections.
}
\end{figure}

The general design of the pipeline is modular. There are separate
programs for each step of the reductions, controlled by a shell script.
Each program can be customized using an extensive list of input parameters.
This provides a relatively easy setup for modifications.

Processing of large 2k$\times$8k frames is done after subdivision into
512$\times$128 pixel subframes with 14 pixel margin to ensure smooth
transitions between solutions for individual pieces. There are 4$\times$64=256
subframes in our case of 2k$\times$8k data. Processing frames piece by piece
makes no real difference for the final photometry except that it enables the
use of the low order polynomials in the modeling of the field distortions and
of the PSF variations. The shape of the small frame reflects the much more
rapid PSF variability in
$Y$ direction compared to the $X$ direction in driftscan images and the fact
that the subtraction algorithm used the same order of PSF variability in both
directions. The first stage of reductions is a construction of the reference
image. A stack of 20 of the best seeing frames with small relative shifts and low
backgrounds is a good choice for the reference frame. The corresponding
shell script ({\tt Make\_template}) takes the list of the images to stack and
determines a crude shift for each of these frames. One of the 20 frames being
stacked together is taken as a coordinate template. All other images will be
resampled to the pixel grid of that image. We used the frame selected
by the OGLE Early Warning System to enforce the agreement of pixel coordinate
systems between our analysis and standard OGLE pipeline.

Then processing of the individual subframes begins. Because of the imperfections in
the telescope pointing we need to find a crude shift between each frame and the
coordinate template. This shift is used to cut the same 512$\times$128 pixel
subframe (with 14 pixel margin) from each of the 20 images. These small images
contain approximately the same piece of the sky. Separate code detects stars in
all subframes and writes ASCII lists to files. Next step is matching star lists
of all images with the coordinate template. A matched ASCII list is created for
each subframe. Another piece of code calculates coordinate transformation and
stores the coefficients in a binary file. The next step is resampling of the
subframes onto a pixel grid of the coordinate template using these
coefficients. These resampled subframes are ready for stacking. The stacking
code takes all current subframes and takes the mean values of the corresponding
pixels adjusted for differential background and intensity scale. This allows us
to renormalize and ``save'' pixels which were bad only on some of the 20 images
and had meaningful values otherwise. In particular, 11 bad columns on the CCD
chip can be totally eliminated in this fashion. All steps of the procedure must
be repeated for each of the 256 pieces of the full format.

The quality of our stacked images is very good. If they are reassembled to form
a single 2k$\times$8k reference frame, there are no discontinuities at the
subframe boundaries. However, for the remaining part of the processing we keep
the reference image subdivided. Still, there may be small differences of the zero
point between the subframes of the reference image due to variable aperture
corrections in the presence of the variable PSF and also due to imperfections
of the derived backgrounds and intensity normalizations. They can be corrected
for later using the overlap regions. Nevertheless linearity is preserved, as
the final value for each pixel is a linear combination of the individual pixel
values with some background offset.

Although the PSF is not required in order to obtain the PSF matching kernel and
the difference frames, we still need it if we want to perform the profile
photometry on the difference images. For each 512$\times$128 pixel subframe we
find spatially variable PSF and store the coefficients in a binary file (Section~3.7).

At this point, with the reference image and its PSF constructed, the main
part of the reductions can be initiated. A separate shell script runs individual
programs for this part. All steps from cutting the same piece of sky
to resampling onto the pixel grid of the coordinate template must be repeated,
this time not only for the 20 best frames, but for all of the data for a given
field. Only after resampling the correct subtraction can take place.
This is the most important part of the processing. Our subtraction code
takes a series of the resampled subframes plus a reference image and determines
the spatially variable convolution kernel, which enables transforming the seeing
of a given piece of the reference image to match that of a corresponding piece
of each test image. Difference subframes are created. Each of them has PSF of
the corresponding test image, but the intensity scale of the reference image.

The difference frames can be measured using profile or aperture photometry,
or both. However before we can measure variables, they need to be found.
Our variable finding code takes a series of subtracted subframes, corresponding
resampled images before subtraction (for noise estimates and mapping of
defects), a reference image, and PSF coefficients for that reference image.
It finds groups of variable pixels, which have the shape of the PSF, and computes their
centroids. Additionally it performs simplistic PSF and aperture photometry
on the reference image at the position of each variable. This crude photometry
does not model the neighboring stars and is sometimes severely contaminated by
the light of the nearby objects. Nevertheless it provides a quick reference
check as to how much flux there is at the location of the variable object.
Coordinates and crude photometry (plus some additional information, see
Section~3.9 for details) are written to a binary file which will be referred to
as ``catalog''. The last step is the actual photometry. The photometric program
makes a single PSF fit to variable light at the location given by the variable
finder. It also performs aperture photometry and determines numerous parameters
of the quality of each photometric point. Section~3.10 contains full details.
Finally it writes the results to a binary file which will be referred to as
``database''.

The following sections give more details on algorithms and implementation.

\subsection{Selection of frames for construction of the reference image}

For each field we need to select the best frames, which will be stacked
together and used as a reference image in all subsequent subtractions.
The properties of these images should be as uniform as possible. After 3 observing
seasons OGLE collected typically 200--300 frames for a bulge field. Among those
about 20 best seeing images also have low background and relative shifts in
the range $\pm 75$ pixels. Therefore we adopted 20 as the number of individual
frames to be coadded. By definition we included the OGLE template image from DoPhot
pipeline (Udalski, Kubiak \& Szyma\'nski 1997) and used it as a coordinate
reference to simplify cross identifications and transformations to celestial
coordinates. All images were carefully inspected visually for possible
background gradients, an occasional meteor, and more importantly bad shape
and spatial dependence of the PSF. About 25 images had to be reviewed before
20 could be satisfactorily included in the reference image. The seeing in the
coadded image was typically 1.1 arcsec while the median for all of our data
is 1.3 arcsec.

\subsection{Shifts between frames}

Before we can track the same piece of the sky in all frames we must first
find a crude shift between each full frame and the coordinate template frame.
This is best accomplished using a cross correlation function

\begin{equation}
\hbox{\it CRF\/}(u,v) = \int f_1(x,y) f_2(x-u,y-v)~dx dy,
\end{equation}

where $f_1, f_2$ are images in question. To find the shift we just need to
find the maximum of this function. For that purpose we used a central
2k$\times$4k piece of each frame. It may seem unnecessarily large, but in this
way the software can tolerate very large shifts, and also there is more signal in the
maximum we are looking for. Such large shifts were the case for several frames,
which otherwise looked normal and had useful pixels in them. This is an
adjustable parameter and could be changed for other applications.

Program {\tt cross} subtracts first a median background estimate from both
frames to avoid excessive base line level for the resulting cross correlation.
Then the images are binned by a factor 16 in both directions to speed
the whole process up. Fourier transforms of both images are taken and cross
correlation function is calculated as

\begin{equation}
\hbox{\it FFT}^{-1}(\hbox{\it FFT}(f_1)\times \hbox{\it FFT}^{*}(f_2)),
\end{equation}

where $*$ indicates complex conjugate. The maximum of that function cannot
be missed, especially in a crowded field! It is sufficient to take the brightest pixel
to be the location of the maximum. Due to binning, the accuracy of this first
guess is 16 pixels. The result will be refined to 1 pixel accuracy using the same
method, but now on 128$\times$128 pixel central piece, adjusted for the initial
guess and without binning. To save time {\tt cross} can calculate
$\hbox{\it FFT}$ of the
coordinate template frame once for the entire series of frames, and write all
shifts to a single file.

\subsection{Detection of stars and centroiding}

For the purpose of detection of point sources the PSF can be approximated
with a Gaussian of some typical width. We take the FWHM to be 2.5 pixels, about
1 arcsec. Program {\tt sfind} calculates the correlation coefficient with this
approximate PSF model at each pixel by convolving with the lowered Gaussian filter
and renormalizing with the model norm and local noise estimate. Convolved
image has pixel values in [0,1] range. Local maxima of this image (defined
by the brightest pixel in a square neighborhood of $\pm4$ pixels) with the
correlation coefficient above 0.7 are added to the list of candidate stars. Objects
with saturated or dead pixels are ignored. In addition, the program outputs
primitive aperture photometry using 1.5 pixel aperture radius and the median
background estimate within an annulus between radii of 3.0 and 7.0 pixels. This
photometry is used only to sort the sources in the order of increasing brightness.
Detection threshold is set to provide $\sim100$ stars for fitting the
coordinate transformation.

Centroids of detected stars are calculated using a 3$\times$3 pixel neighborhood
centered on the brightest pixel of a given star. To obtain the centroid
in say $X$ we integrate the flux in such neighborhood along $Y$ to get
3 flux values at 3 integer values of $X$, and then find the location
of the maximum of so defined parabola. We repeat this for $Y$. This
simple algorithm for our particular purpose gave consistently better accuracy
than any other prescription we tried, e.g. fitting the position using a
Gaussian approximation to the PSF.

\subsection{Cross identification of stars between images}

Cross identification of star lists for two images is done using a variation
of the triangle method (program {\tt xymatch}). It does not require
the initial tie information, although our subframes are already corrected for
the crude shift (Section~3.2). Because of the nature of field distortions
in driftscan imaging using OGLE telescope, the local residual shift with
respect to the mean value for the entire 2k$\times$8k format can reach several
pixels. The algorithm starts with the lists of all triangles that can be formed
from stars in both images. A triangle is defined by: the length of the longest side,
the ratio of the longest to shortest sides and the cosine of the angle
between those sides. In so defined 3 dimensional space the program looks for
close points using a combination of fractional and absolute tolerance levels.
Because the cost of this method is ${\sim}n\times (n-1)\times (n-2)/6$ we can
afford only 20--30 stars for the initial matching. These are selected to be
the brightest stars in both lists, therefore the size of the subframe cannot
be too large. In the case of a large format the slightest difference in the stellar
magnitude corresponding to the saturation level, e.g. due to seeing variations,
would shift the tip of the luminosity function by much more than 20--30 stars,
making both lists exclusive. The initially matched list of $\sim20$ stars
provides the linear fit to the coordinate transformation. It is sufficient for
identification of all remaining stars needed in the final fit.

\subsection{Resampling pixel grids}

Program {\tt xygrid} takes the matched lists of coordinates for stars
in two images and fits the full polynomial transformation between the two
coordinate systems, which will enable the difference in the field distortions
to be taken out. For our 540$\times$156 pixel subframes we use 2nd order
polynomials. The fit is cleaned by the iterative rejection of the points deviating
by more than 3$\sigma$ from the current best fit. Typical scatter in the
matched positions is 0.06 pixels, consistent with our discussion of centroid
errors in Section~4.2. It can be safely assumed that
transformation is accurate to 0.1 pixels. Coefficients stored in a binary file
are then used in program {\tt resample} to interpolate a given subframe.
We use a bicubic spline interpolator (Press et al. 1992). Pixels for which
there is no information are given values which will be later recognized
as saturated. At this point the images can be subtracted or coadded.

\subsection{Image coaddition}

Preparation of the reference image requires stacking of frames, which is
a relatively simple problem, since it does not require matching of the PSFs.
If it were not for saturated pixels, bad columns and edge effects
due to shifts, after resampling a simple mean value of each pixel would
suffice. Things change if we want to save pixels which are bad only on some
of the stacked images, but otherwise have photometric information in them.
We need to adjust for different background and scale of each frame, at least
to the level when the effects of the patched defects in the final result are
negligible.

Our simple algorithm for stacking was implemented in program {\tt mstack}.
We start with a series of 20 subframes. The first of them is a piece of
the coordinate template and will also become the reference for background
and scale. For each frame we prepare a histogram of pixel values.
It is dominated by a broad sky peak at low values, followed by a much weaker
wing due to stars, which extends all the way to the saturation level.
In a crowded field the sky peak is heavily skewed by faint stars. We found that
the simple fit for the scale and the background difference to all the pixels
in the image is very unreliable. The results are sufficiently accurate if we
take the part of the peak inside its own FWHM and consider the flux such that
30\% of the pixels in this trimmed distribution lies below. Then, for bright
pixels we consider the ratio of their values in each image to the value in
the first image (with the backgrounds subtracted). Assuming that the PSFs of all
frames going into the reference image are similar, the variance weighted mean
of the pixel by pixel ratios is an estimate of the scaling factor. The assumption
is justified by the narrow range of the seeing allowed for the frames which were
used in the construction of the reference image. To assure that the compared pixels
belong to stars the minimum pixel level for this comparison is 300 counts
above the upper boundary of the FWHM region of the sky peak. The results
of these renormalizations are very good. It is impossible to tell which areas
of the final frame have been recovered from minor bad spots. In particular
the strip of 11 bad columns on the CCD chip of the OGLE camera disappears
completely. It is important to realize that even if the backgrounds
and scaling factors were in error, the pixel value of the final combined image
would still be a linear function of the individual pixel levels (with the background
offset). This matters only for noise estimates, but in our case the reference
image constructed here is treated as noiseless. Noise estimates later on are
taken from individual frames.

The program has an option of using the median statistic instead of the mean.
However it should not be used unless the seeings of the frames are matched.
In this case even slight problems with the background levels and scalings
will result in significant nonlinearity.

\subsection{PSF calculation}

As mentioned before, the PSF is not required in order to obtain the PSF
matching kernel and the difference frames. We determine the PSF solely
for the purpose of the profile photometry on the difference images.

Substantial part of program {\tt getpsf} deals with the selection of good PSF
stars. First the full list of candidate objects is selected at local maxima
of the intensity, for which the highest pixel stands out by more than
2$\sigma$ above the background, where $\sigma$ is the photon noise estimate.
A simplistic value
of the flux is calculated using an aperture with 3.0 pixel radius. The frame
is subdivided into 64$\times$64 pixel boxes to ensure the uniform density of
candidate PSF stars by selecting approximately the same number of stars in
each box. We require about 100 PSF stars for the fit taken from bright end of
the luminosity function. The peak value for a star is refined using parabolic
fits in the 3$\times$3 pixel area around the central pixel. The ratio of the
background subtracted peak to the total flux in the object measures light
concentration and is required to be less than 20\% for stars. For most cosmic
rays this parameter is much larger. The sample of well behaved stars is cleaned
of misshapen objects, e.g. cosmic rays and very tight blends, using
sigma clipping on the distribution of light concentrations. Finally candidates for
the fit are checked for close neighbors. The star is rejected from the fit if
in the area $\pm 3$ pixels around the peak there is another local maximum at
least 2$\sigma$ above the background and brighter than
$0.15\times r\times f_{\rm peak}$, where $r$ is the distance in pixels and
$f_{\rm peak}$ is the peak flux of the candidate star. With typical FWHM seeing
values of around 3 pixels, this eliminates stars which have fluxes
significantly contaminated due to crowding.

The PSF model consists of two Gaussians, one for the core and one a factor
of 1.83 wider for the wings, each multiplied by a 3rd order polynomial.
Both Gaussians are elliptical. The position angle of the major axis and
ellipticity can vary but remain the same for the core and wing components.
Spatial variability is modeled by allowing each of the local polynomial
coefficients to be a function of $X, Y$ coordinates across the format, also
a polynomial, but this time of the 2nd order. The above procedure is similar
to the algorithm for fitting the PSF matching kernel in Section~3.8, for which
there are published descriptions (Alard \& Lupton 1998 and Alard 2000).

The first guess for the shape of the Gaussians is taken to be circularly symmetric
and the initial FWHM of the core component is 3.0 pixels. The linear and
nonlinear parts of the fit are separated. The shape of the Gaussian and its centering
are nonlinear parameters and they are adjusted iteratively since the required
correction is often minute, certainly for our data. Also, the individual
amplitudes of stars must be taken out before the fit to generic PSF parameters.
Therefore in each iteration we first solve the linear problem for all
polynomial coefficients, then we update the shape of the Gaussian using moments
of the light distribution of the current fit and finally we correct centroids
of fitted objects with linearized least squares and recalculate the norm
of each star. To avoid any potential instability the first few iterations are
done with spatial variability turned off and then the full fit can be safely
completed. In our case only 2 iterations are required at each stage.

PSF coefficients are stored in a binary file for later use in detection of
variables and profile photometry on difference images.

\subsection{Subtraction}

The goal at this stage of the data processing is to find the best PSF matching
kernel and subtract the reference image from all the remaining images.
A detailed description of the method for optimal image subtraction is given by
Alard \& Lupton (1998). Alard (2000) presents a very refined algorithm with
spatially variable kernels and flux conservation. Here we describe our
implementation and parameters used with the OGLE bulge microlensing data. The
corresponding program is called {\tt aga}. It takes a series of frames
resampled onto a common pixel grid of the reference frame. We did not use the
capability for external masking of unwanted pixels, because the internal
rejection algorithms gave us satisfactory results.

The heart of the method is the choice of the kernel decomposition: 3 Gaussians
of constant widths multiplied by polynomials. The kernel in this form is linear
in the parameters, making the solution of Equation~2 simply a big least square
problem. We used Gaussians with sigmas $\sigma$ = 0.78, 1.35, and 2.34 pixels
modified by 2D polynomials of orders $n$ = 4, 3, and 2 respectively. The above
parameters previously gave us good results for ground based data sampled near
the Nyquist frequency (Wo\'zniak et al. 2000, Olech et al. 1999). Convolutions
are performed directly, i.e. in real space, using a 15$\times$15 pixel rasters
for the kernel components. This is considerably faster than Fourier calculation
in the case of a large difference between the spatial scales of the functions
to be convolved. Spatial variability is introduced by allowing each of the
local kernel coefficients to be a function of $X, Y$, again a polynomial,
to keep things linear. The spatially variable problem quickly grows, the number
of coefficients is $(n_{\rm spatial}+1)\times(n_{\rm spatial}+2)/2$~~ times
larger than in the case of a constant kernel and can exceed 100 in normal
applications, still affordable in terms of the S/N ratio given the enormous number
of pixels. We used second order spatial dependence, $n_{\rm spatial}=2$.
The program also fits the difference in backgrounds between the images. A first
order polynomial was used for that purpose.

The first step is convolving a reference image with each piece of the kernel to
form a set of images, which can be viewed as the basis vectors. Solving
Equation~2 means finding a linear combination of these basis vectors that
closest reproduces the light distribution of the frame to be differenced.
Because of the computing time requirements spatial variability is handled by
subdividing the fitted area into a number of square domains, sufficiently small
so that the PSF variability can be ignored inside a single domain. Local
kernel coefficients at the domain center are adopted for the entire domain,
here 23$\times$23 pixels (see Alard 2000 for detailed derivations). Some pixels
are better left unused, e.g. those which vary because they belong to variable
stars, not due to seeing. Also one should avoid fitting large areas dominated
by the background, where all the structure is dominated by noise and the resulting
kernel will display the large amplitude, high frequency oscillations. In crowded
fields this is never a problem. Due to the finite width of the kernel for some
pixels the value of the convolution cannot be determined. Pixels near the edges
of the image or near unusable pixels are rejected with the safety margin of 7 pixels
(half width of the kernel) for convolutions and 2 pixels for other images. Two choices of
the domain patterns are available: uniformly distributed domains or centered on
bright stars. By trial and error we determined that the kernel fits are best
in the second mode with 20$\times$10 individual domains spread over the area
of a subframe. Once the appropriate domains in the basis images have been selected,
we get the first guess for the solution. To solve the Equation~2 we used
LU decomposition from numerical recipes throughout our programs (Press et al. 1992).
The initial solution is cleaned with sigma clipping of individual
pixels within domains and clipping of the entire distribution of whole domains
using their $\chi^2$ per pixel values. We require that after sigma clipping
at least 75\% of the domain area must be acceptable for the domain to enter
the final solution. Also at least 40\% of the total fitted area must be left
in the fit and the final $\chi^2$ per pixel must be less than 8.0 for
the program to declare a successful subtraction. If this is the case, the reference 
image is convolved with the best fit kernel and subtracted from the program subframe.
Otherwise the subframe is rejected and flagged as such. At the end the code
writes the difference frames and kernel coefficients to binary files.

It turns out that for the OGLE-II bulge data the solution is dominated by red
clump stars with $I\/\simeq15.5$ mag. The photometric errors for these bright
stars are already influenced by systematic effects of seeing and PSF uncertainties
due to the resolved background. To acknowledge the sources of error other than
just the photon noise we rescaled the photon noise estimate by a factor of 1.6,
which resulted in average $\chi^2$ per pixel = 1 in the difference images. We used
this fudge factor in selection of variables and for error bars in the database
light curves. After we had completed our reductions, we discovered that this was
actually a pretty big overestimate for faint stars (see Section~4.1).

\subsection{Finding variables and centroiding}

We decided to detect variable objects using some preliminary variability
measures based on the entire series of difference images for a given field,
and make final measurements only for those candidates. We also avoided
strong assumptions about the type of flux variations to be extracted.
The main idea is to encode all interesting variability of several basic types
in a corresponding number of ``variability images'', find variables in these
frames and calculate their centroids.
A single value for the centroid for each variable is calculated
using the entire series of difference frames, which eliminates the need for
cross identification of variables between images and enables the measuring of
photometric points for frames on which the difference signal for a given variable
has not been detected. This way our databases contain only the light curves for
candidate variables, non-variable stars will not be included. Our algorithm for
finding variables may not seem especially natural, but it is the most
efficient we could find in the sense that it recovers practically all stars which
appear variable upon visual inspection of the difference frames, and does not
return too many spurious detections. In fact about 80\% of the 4597 candidate variables
in the database for SC1 bulge field could be classified as one of the known types of the
periodic variables or had significant night-to-night correlations in their
light curves, which are a strong indicator of real variability (Mizerski \& Bejger,
private communication). The remaining objects are either non-variable stars
which passed our selection cuts, or ghost variables caused by various undetected
problems, e.g. the telescope tracking errors or cosmic rays. We deliberately
admit some noise background in the catalog to provide the testing ground for new
automated variability classification schemes.

Program {\tt getvar} starts by rejecting some fraction of the frames with
the worst seeing, in our case 10\%. It also uses a conservative value for the
noise estimate, 1.6 times photon noise. This factor matches the average quality
of subtraction measured by $\chi^2$ per pixel and is a correct scaling
for red clump giants, bright stars which dominate the solution of the main
equation for the PSF matching kernel (Equation~2).
For faint stars this is an overestimate, as shown in Section~4.1, but
the detailed noise properties of the data were not known at the time when
parameter values had to be selected.

In the next step we consider individual light curves in the 3$\times$3 pixels
square aperture centered on every pixel. This corresponds to smoothing of all
difference frames with the 3 pixels wide mean filter before examining pixel
light curves. Some points are rejected for saturated and dead pixels and we
require that at least 50\% of the measurements remain in the cleaned light curve.
For each pixel light curve we take the median flux to be the base line
flux and analyze the ratios of the departures from this base level to their
noise estimates. The specific numbers we quote here are all adjustable
parameters of the program. To include periodic and quasi periodic variables
which vary continuously, as well as eclipsing binaries and other transient
phenomena like flares and microlensing, we have two channels for selecting
variable pixels. The pixel is declared as variable if one of the two conditions
are met:

\begin{enumerate}
\item there are at least 3 consecutive points departing at least 3$\sigma$ from
the base line in the same direction (up or down), or
\item there are at least 10 points total departing at least 4$\sigma$ from the
base line in the same direction, not necessarily consecutive.
\end{enumerate}

In the next step we label variable pixels according to the ratio of the number
of the deviating points from the above cut which depart upwards to the number
of points departing downwards.
If the ratio is between 0.5 and 2.0 we fill a corresponding pixel of the
variability image for ``continuous'' variables. Otherwise we fill the pixel
of the image for ``transients''. As the measure of pixel variability we adopt
$D=\sum_i|F_i - F_{\rm 0}|$, where $F_i$ is the flux and
$F_{\rm 0}$ is base line flux. For ``continuous'' variables the pixel value in
the variability image will be:
$(D_{\rm up}+D_{\rm down})/(n_{\rm up}+n_{\rm down})$, where $n$ is the number
of points high or low with respect to $F_{\rm 0}$. For ``transients'' variability
image will contain: $D_{\rm up}/n_{\rm up}$ or $D_{\rm down}/n_{\rm down}$
depending whether $n_{\rm up}>=n_{\rm down}$ or $n_{\rm up}<n_{\rm down}$
respectively.

After the variability images are constructed we can look for groups of variable
pixels. With the above definition of $D$ sufficiently high signal to noise
variables will produce groups of pixels resembling the local PSF shape.
Therefore the last step is to detect ``point sources'' in variability images
using a PSF model, a very similar procedure to our star detection algorithm
(Section~3.3).
Just like in the case of star detection we look for local maxima of the
correlation coefficient with the PSF in the excess of 0.7. In the end we have
two lists of candidate variables of the basic types we described.

As mentioned before we determine the centroid of a variable only once using
entire series of difference images. The program takes 9$\times$9 pixel rasters
centered on each variable from frames on which the difference between the
measured flux and the template flux was at least 3$\sigma$. The absolute values
of these rasters are subsequently weighted by their signal to noise and coadded
to accumulate as much signal in the peak as possible. Using a 3$\times$3
neighborhood of this peak and parabolic fit we calculate the centroid in
exactly the same way as for regular frames (Section~3.3).

Cross identification of variables from ``continuous'' and ``transient''
channels is done because some of them can in principle appear in both lists.
Variables closer than 2.0 pixels are treated as one and are given the value
of variability type 11. ``Transients'' are type 1 and ``continuous variables''
are type 10. Currently these are all variability types included, but the extension
to other types should be straightforward. One interesting example to consider
in the future might be a single high point, which would filter moving objects.

The final step is simple PSF and aperture photometry on the reference image at
the location of the variable. It must be emphasized that this photometry does
not attempt any modeling of the surrounding stars and therefore for faint
and/or blended variables it can be severely contaminated by the neighbors.
This information provides only a quick check of how much flux there
is in the template image at the location of the variable, because the actual
light curve contains only the difference signal. We also set a crowding flag
equal to 1 if there is a pixel brighter than $0.15\times r\times f$ in the
$\pm 4$ pixels neighborhood, where $r$ is the distance (in pixels) and $f$ is
the flux of the central pixel of the variable on the reference image. Given that
in the reference images the FWHM of the seeing disk is typically less than 3 pixels,
for an object with the crowding flag set to 0 (uncrowded) it will be likely that
less than 10\% of the flux within its PSF belongs to the neighboring stars.

Program {\tt getvar} writes a catalog entry for each of the variables.
The format of the 52 byte record is the following (all fields are 4 byte
{\tt float} numbers except for the last 4, which are of 4 byte {\tt integer}
type; the most significant byte is stored first):

\begin{tabular}{rl}

1. & $X$ template coordinate \\
2. & $Y$ template coordinate \\
3. & flux -- profile photometry \\
4. & flux error -- profile photometry \\
5. & flux -- aperture photometry \\
6. & flux error -- aperture photometry \\
7. & background \\
8. & $\chi^2$ per pixel for the PSF fit (usually bad fit due to neighboring
stars)\\
9. & correlation coefficient with the PSF \\
10. & number of bad pixels \\
11. & variability type \\
12. & number of frames used for centroid calculation \\
13. & crowding flag \\

\end{tabular}

With this paper we make a public release of the pipeline output for the first
OGLE-II bulge field (SC1). To facilitate the easiest possible data access we decided
to format the distribution files as ASCII tables. Section~6.2 provides the details.

\subsection{Photometry}

We perform both profile and aperture photometry on our difference images,
keeping the centroid fixed. Aperture photometry and its noise are given
by equations~5 through 7.

\begin{equation}
a_{\rm ap}=\sum_i^{r_i<r_{\rm ap}} f_i
\end{equation}

\begin{equation}
\sigma_{\rm ap}=\root \of{\sum_i \sigma^2_i}
\end{equation}

\begin{equation}
\sigma^2_i={f_{i, 0}\over G},
\end{equation}

where $f_i$ is the difference flux in pixel $i$,
$f_{\rm i, 0}$ is the actual pixel flux including background and before
subtraction of the reference image, the sum is over pixels with centers within
the aperture radius $r_{\rm ap}$ from the centroid. $G$ is the gain in
$e^-$/ADU.

Profile photometry comes down to a 1 parameter fit for the amplitude
with

\begin{equation}
\chi^2 = \sum_i^{r_i<r_{\rm fit}}
{(a_{\rm psf} P_i-f_i)^2 \over \sigma^2_i}.
\end{equation}

$P_i$ at pixel $i$ is the value of the PSF profile centered on the
variable and the sum is within the fitting radius $r_{\rm fit}$ around
the centroid. The best fit is given by:

\begin{equation}
a_{\rm psf}= {
\sum_i (f_iP_i/\sigma^2_i) \over \sum_i (P^2_i/\sigma^2_i)}
\end{equation}

\begin{equation}
\sigma_{\rm psf}= {1 \over \root \of {\sum_i (P^2_i/\sigma^2_i)}}
\end{equation}

Obviously the PSF photometry gives optimal noise and allows the meaningful
renormalization for the rejected saturated and dead pixels. Program {\tt phot}
takes a series of difference frames, resampled frames before subtraction for
noise estimates, coefficients of spatially variable PSF matching kernel
for each subframe and finally coefficients of the PSF for the reference frame.
By convolving local kernel and reference PSF for each subframe it constructs
a PSF profile at the position of each variable and calculates amplitudes
$a_{\rm ap}$ and $a_{\rm psf}$ with corresponding errors $\sigma_{\rm ap}$
and $\sigma_{\rm psf}$, which involves little more than simple sums over
pixels. We used $r_{\rm ap}=r_{\rm fit}=3.0$ pixels. Whenever there is
no information {\tt phot} will put requested error codes to keep the record
of such gaps. The final light curves are stored in a binary file. Records for
all epochs for a given variable must be written before the next light curve can
be stored. The total number of 40 byte records is equal to
$n_{\rm variables}\times n_{\rm epochs}$. The time vector is the same for all
stars and therefore it is more efficient in terms of storage to keep it separately
in a short ASCII file.
All fields of the record are 4 byte {\tt float} numbers except for the last
one, which is a 4 byte {\tt integer}; the most significant byte is stored first.
The format of the binary record is the following:

\begin{tabular}{rl}

1. & flux -- profile photometry \\
2. & flux error -- profile photometry \\
3. & flux -- aperture photometry \\
4. & flux error -- aperture photometry \\
5. & background \\
6. & $\chi^2$ per pixel for the PSF fit \\
7. & correlation coefficient with the PSF \\
8. & $\chi^2$ per pixel of subtraction for entire corresponding subframe \\
9. & FWHM of the PSF profile \\
10. & number of bad pixels within the fitting radius \\

\end{tabular}

As in the case of the catalog the results for SC1 field in public domain
are available in ASCII format (see Section~6.2 for details).

\section{Performance}

\begin{figure}
\plotfiddle{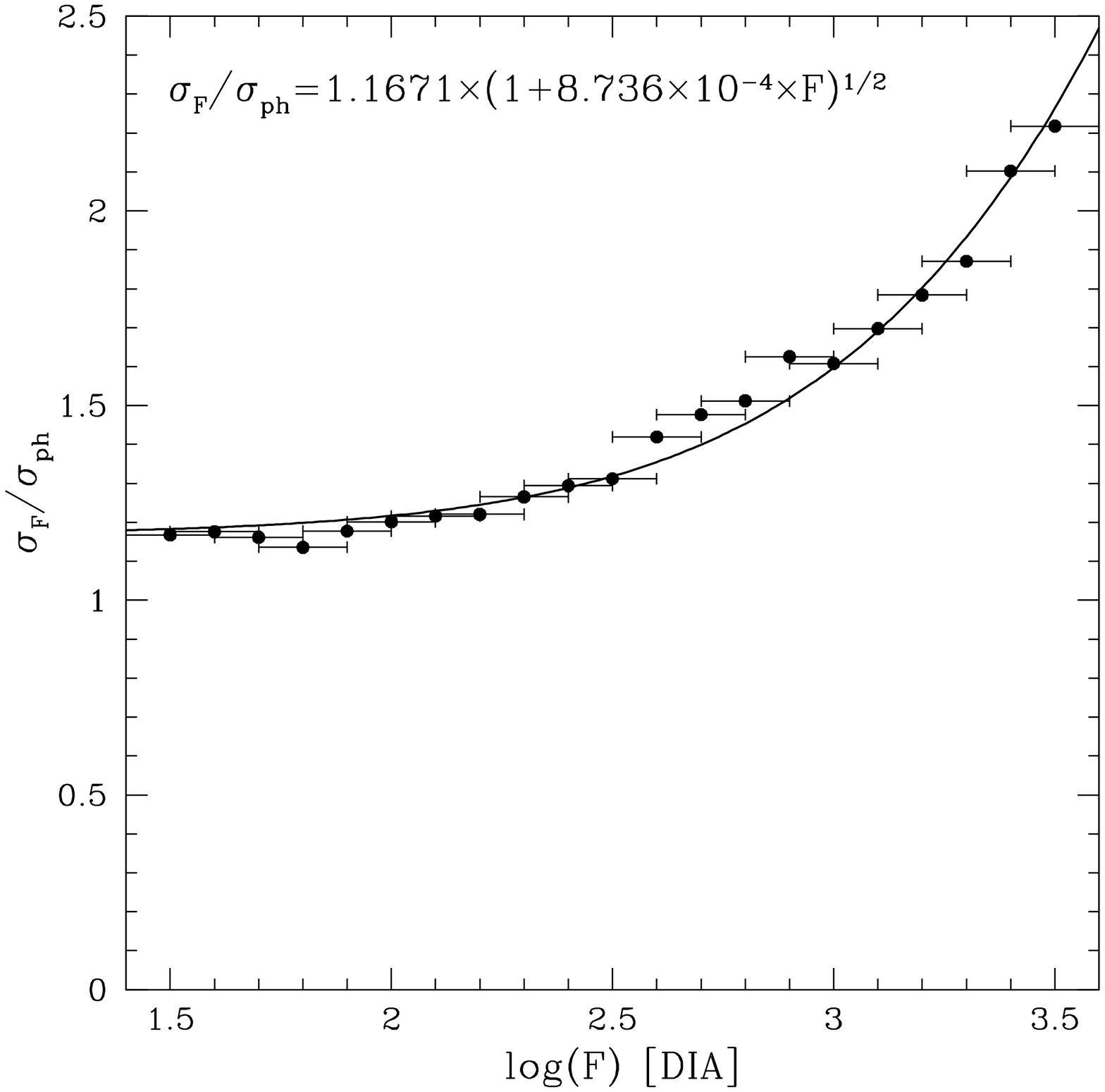}{8cm}{0.}{70.}{70.}{-225}{-100}
\caption{The ratio of the actual scatter to the photon noise estimate as a
function of flux logarithm. The data in this plot comes from 483 classic single
microlensing events. Only points outside the $t_{\rm max}\pm 2 t_{\rm 0}$ region
of each event were used to get residuals around the best fit model (Section~4.1).
Error bars indicate the size of the flux bin. The solid line
is the empirical fit to the data points:
${\sigma_F/\sigma_{\rm ph}}=1.1671~\root\of{1+8.736\times 10^{-4} F}$.
}
\end{figure}

\begin{figure}
\plotfiddle{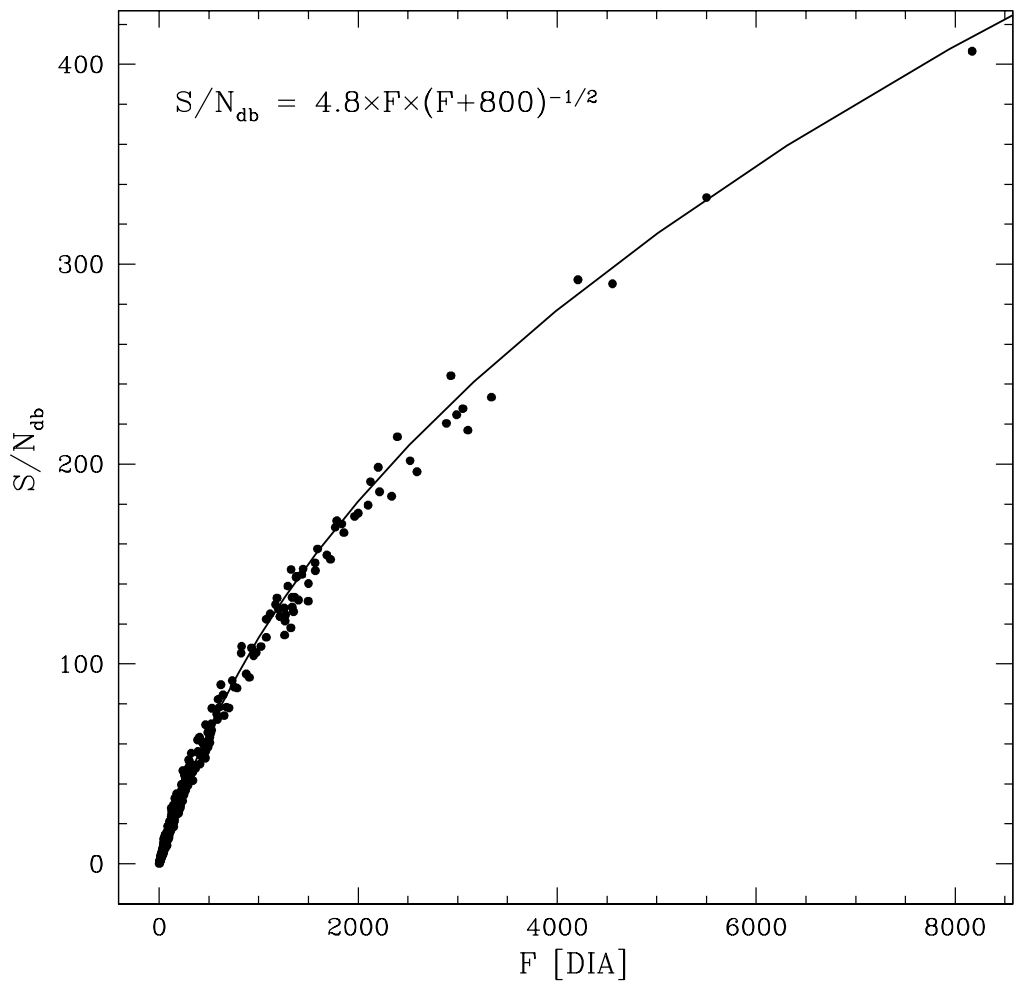}{8cm}{0.}{70.}{70.}{-220}{-100}
\caption{Nominal signal to noise in our DIA database as a function of flux for
483 microlensing events. Photon noise was multiplied by 1.6 in the DIA
database and in the figure. The solid line is the empirical fit to the data
points: ${\rm S/N_{db}}=4.8~{F/\root\of{F+800}}$. This relation combined
with the one in Figure~3 gives the actual scatter as a function of flux.
}
\end{figure}

\begin{figure}
\plotfiddle{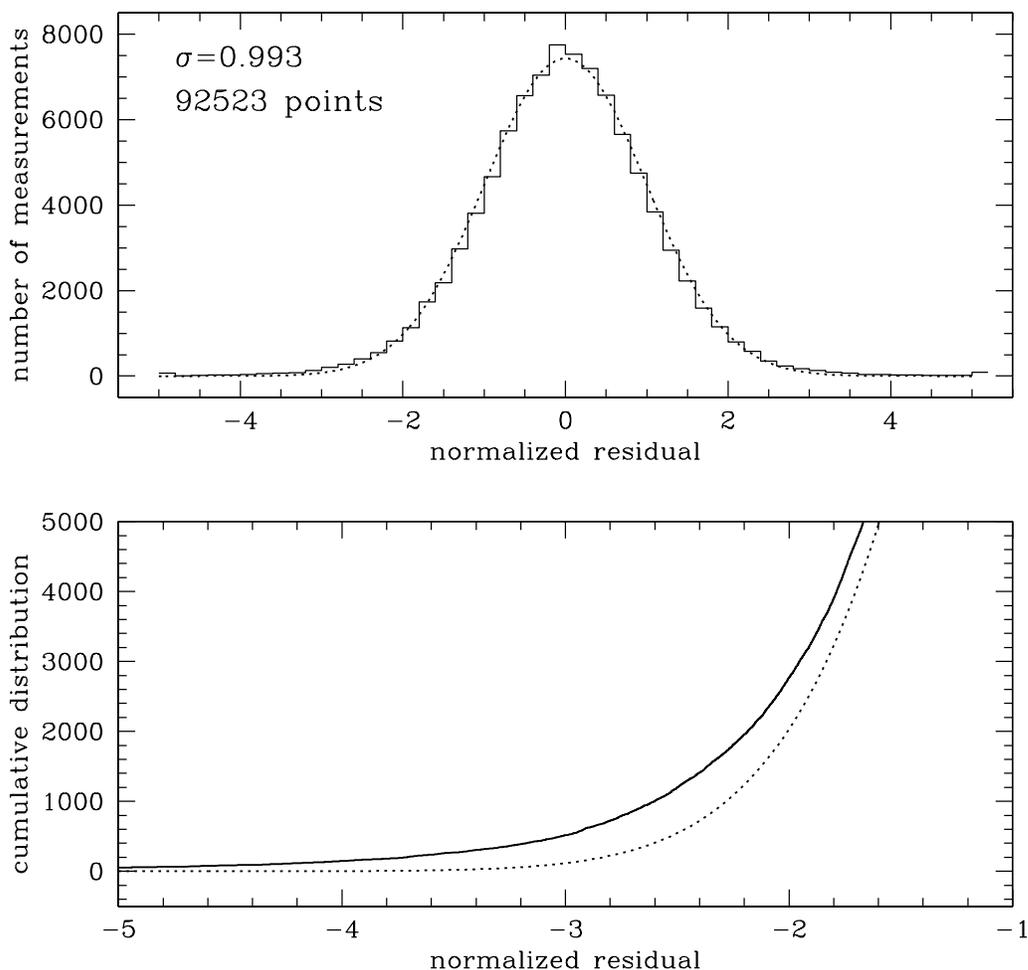}{9cm}{0.}{70.}{70.}{-225}{-100}
\caption{Differential (upper) and cumulative (lower) distributions of
normalized residuals for 92523 individual measurements of all microlensing
events in Figure~3. Each residual was normalized by its own error
estimate. For the error bars we adopted photon noise corrected using the fit in
Figure~3. The solid lines are the data histograms and the dotted lines
indicate Gaussian distributions which have the same centered range containing
68.3\% of the points and the same norm as the data histograms. Please note that
in the lower panel we showed only the left wing of the cumulative distribution.
The full distribution spans the range between 0 and 92523.
The non-Gaussian tail amounts to about 700 points, 0.7\% of the total.
}
\end{figure}

\begin{figure}
\plotfiddle{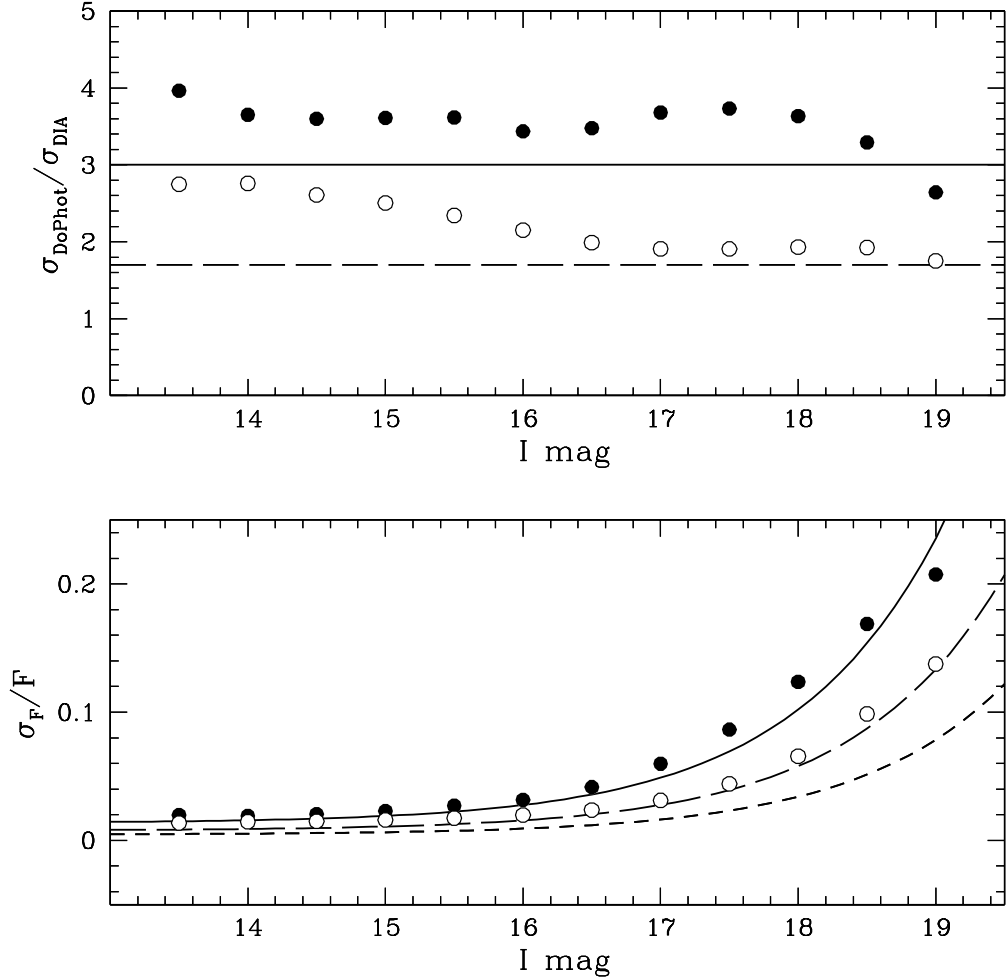}{11cm}{0.}{70.}{70.}{-225}{-100}
\caption{Fractional flux error as a function of magnitude for OGLE bulge
catalogs compared to the results from difference image photometry. In the lower
panel we show the median scatter in 12 magnitude bins after removing 10\%
of the high points to allow for variables. The solid points are the data for
SC4, a very dense field, and the open circles for SC10, a relatively
sparse field. The short dashed line indicates the scatter in difference image
photometry (the fit from Figure~3). The long dashed and solid lines are the
same as the short dashed line, but multiplied by 1.7 and 3.0 respectively.
The upper panel shows the improvement factors: the same data as in the lower
panel, divided by the short dashed line. The falling trend in the improvement
factors towards the faint stars is artificial, and comes from the fact that
the selection criteria for frames used in the OGLE catalog entries are tighter
for fainter stars.
}
\end{figure}

\begin{figure}
\plotfiddle{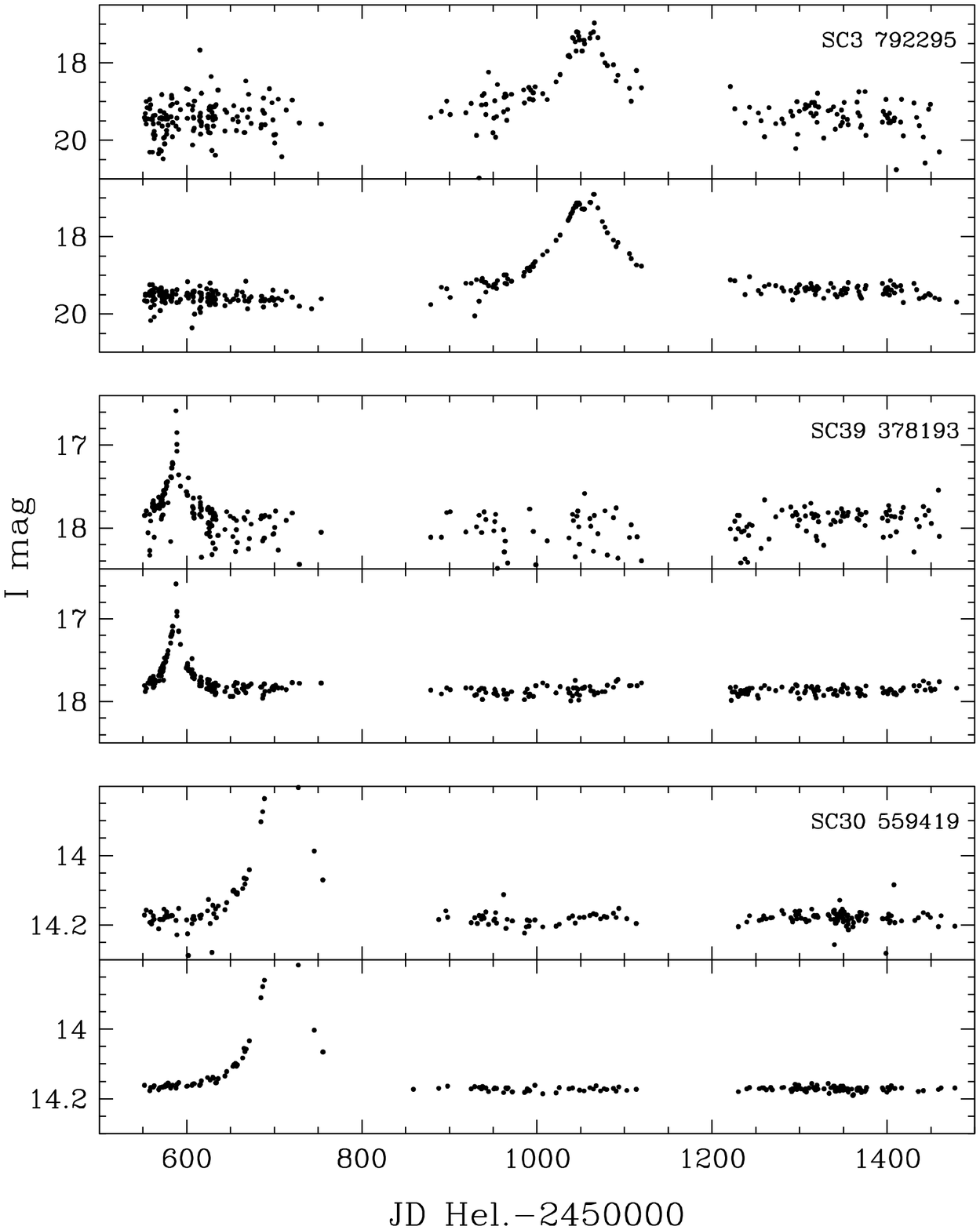}{15cm}{0.}{70.}{70.}{-215}{-10}
\caption{Comparison of the difference image photometry with DoPhot output
in the OGLE database for three sample microlensing events. Each of the three
panels shows two light curves of a very faint, average and bright star,
from top to bottom. Despite the fact that our light curves typically contain
a few more points (we allowed even the worst seeing frames), image subtraction
provides dramatically reduced scatter and improved information about events.
}
\end{figure}

\subsection{Noise characteristics}

Ideally one would test the properties of the photometry using a sample of
constant stars. However our catalogs contain only suspected variables. Some of
the bright non-variable stars are included in the catalog because at the time
the variables were selected we did not know the exact behavior of the noise,
but they are not typical. The next best stars are microlensed stars:
these have a long base line, when the light is essentially constant and the form
of light variation is known in typical cases. The full catalog of microlensing
events from this analysis will be published shortly. Here we will only make the
tantalizing statement that 512 events were found in all 49 fields, about
a factor of 2 increase compared to the OGLE catalog (Udalski et al. 2000) with
the appropriate adjustments for differing selection criteria. Perhaps
even more significant is the fact that finally a sample of 305 well covered
events with good S/N could be fully algorithmically extracted from the OGLE data
without any unwanted ghost light curves coming through the selection process.
The MACHO team recorded a similar fractional increase in the number
of events detectable using their own algorithm (Alcock et al. 1999b).

Noise properties were derived from the residuals of individual measurements
around the best fit microlensing curve. Points inside the region
$t_{\rm max} \pm 2 t_0$ were rejected ($t_{\rm max}$ is the moment of maximum
light and
$t_0$ is the Einstein radius crossing time). Also events with evidence for
systematic departures from classic point source microlensing curve and evidence
of source variability were excluded from the analysis. Each residual was normalized
by the photon noise estimate for the corresponding photometric point. Then stars
were grouped according to the total flux on the template at their location
filtered through the PSF (the one in the catalog entry, Section~3.9).
All residuals coming from light curves of stars in a given group were
merged into one distribution.

For each group we calculated the half width of the region containing 68.3\% of
residuals and centered on 0, a robust estimator of the width of the Gaussian
distribution. In Figure~3 we plot thus estimated $\sigma$
as a function of flux. A red clump star at $I\simeq15.5$ mag has the flux
around 1300 counts in our units. The width of the log flux bin corresponds to 
0.5 mag, as indicated by horizontal error bars. At the faint end we find that the
noise (standard deviation) is only 17\% above the Poisson limit. The data is
well fitted by the following simple law:

\begin{equation}
{\sigma_F\over\sigma_{\rm ph}}=1.1671~\root\of{1+8.736\times 10^{-4} F}.
\end{equation}

As we move towards brighter stars this excess increases due to systematic
effects related to seeing and PSF uncertainty,
first very slowly and then faster. Alard and Lupton (1998) provide a possible
explanation in terms of the atmospheric turbulence. In Figure~4 we also show the
error estimate given in our databases, as a function of flux. This is
basically photon noise multiplied by 1.6 and propagated through the PSF fit.
The formula

\begin{equation}
{\rm S/N_{db}}=4.8~{F\over\root\of{F+800}}
\end{equation}

provides a good fit to the data. Therefore the actual S/N ratio is:

\begin{equation}
{\rm S/N = 1.6~{\left(\sigma_F\over\sigma_{ph}\right)}^{-1} S/N_{db}}.
\end{equation}

The solid curve shown in Figure~3 can be used to rescale the error bars
in order to improve the consistency of the light curves.
Then we may also consider renormalized residuals with this
new rescaled noise and check their integrated distribution, that is all
points from stars of all magnitudes. The resulting histogram is shown in
Figure~5 along with the Gaussian distribution of the same norm and $\sigma$
estimated from the 68th percentile of the histogram. The departures from
gaussianity are very small. The cumulative distributions in the lower
panel of Figure~5 reveal a non-Gaussian tail at the level of only $\sim0.7$\%.
Regular, understandable behavior of the noise will certainly have a positive
impact on the amount and quality of the information derived from these data.

To further evaluate the difference image photometry we compared the results
of Figures~3 and 4 to the photometry obtained with DoPhot in the standard OGLE
pipeline. To simplify our task, we used stars from the OGLE bulge catalogs.
Each OGLE catalog entry contains the information about the mean and the scatter
of the so called ``good'' measurements, selected using the values of seeing and
of the error bar returned by DoPhot compared to the typical errors for stars
in a field of a given density (Udalski et al. 1993,
Szyma\'nski \& Udalski 1992).
We considered stars in 12 magnitude bins between
$I$=13 and 19 mag detected in two fields: SC4, a dense field,
and SC10, a relatively sparse field. In SC4 and SC10 on the best frame
DoPhot detects about 770,000 and 460,000 stars respectively. In each
magnitude bin we reject 10\% of the stars with the largest scatter,
a conservative allowance for real variability. In the remaining group of stars
we adopt median scatter as the measure of typical behavior. The results are
summarized in Figure~6. The points indicate the actual median scatter
from OGLE database. We also show three lines which correspond to the scatter
in image subtraction method (short dashed) and its two copies rescaled by
factors 1.7 (long dashed) and 3.0 (solid).
In the entire range of fluxes difference image photometry is at least 2 times
more accurate, sometimes almost 4 times. The fact that the improvement seems
to decline for faint stars is an artifact of the cleaning procedure used
by OGLE. Figure~7 is a nice illustration of what can be accomplished. For each
of the three sample microlensing events with sources of differing brightness
we show two light curves. One obtained with image subtraction software and one
with DoPhot. The improvement is striking. In the case of star SC3 792295 the
binary nature of the event is obvious in the Difference Image Analysis,
while with DoPhot photometry alone one would have to accept a large uncertainty
margin for such hypothesis.

\subsection{Accuracy of the centroid}

\begin{figure}
\plotfiddle{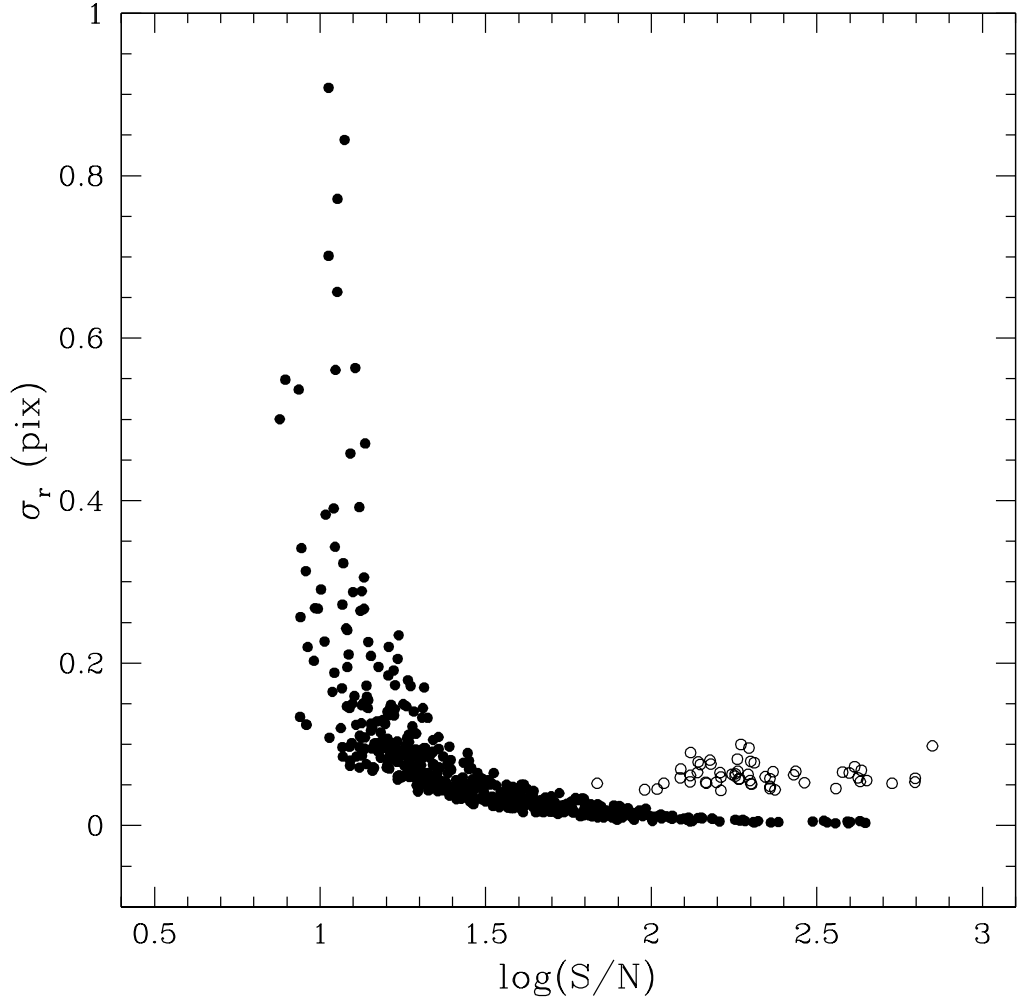}{9cm}{0.}{70.}{70.}{-225}{-100}
\caption{Quality of the centroid as a function of S/N ratio in the intensity
peak used for the estimate. The solid points represent the scatter measured in
100 simulated realizations of a 3$\times$3 neighborhood of the intensity peak.
The open points correspond to the actual scatter for bright isolated stars
measured in 100 frames. At the bright end the atmosphere limits the accuracy to
about 0.06--0.10 pixels (0.024--0.040 arcsec), a much larger value than
suggested by the simulation. Near S/N=15 and below centroid suddenly becomes
uncertain.
}
\end{figure}

To complete the discussion of how the codes perform we summarize the properties
of the centroid. The expectation is that the random error in position of the
intensity peak depends on the width of the peak (curvature), the signal to
noise ratio and the exact centroid position with respect to the center of the
pixel. This is independent of whether the peak is a normal image of a
star or was obtained by coadding peaks from variables on difference frames,
as in the procedure of centroid finding for variables in our catalog
(Section~3.9). In Figure~8 we compare two estimates. The first estimate comes
from a simulation (solid points). We generated fake 3$\times$3 pixel peak
neighborhoods of our microlensing events using their actual catalog centroids.
The photon noise was included in the simulation and normalized to the total
S/N ratio in the peak from which the catalog centroid was calculated. Figure~8
shows the standard deviation of 100 realizations of the experiment using our
standard centroiding procedure of Section~3.9. The range of seeing is
relatively narrow in entire data set so the correlation with S/N clearly shows.
For S/N$>$20 the position is accurate to a small fraction of a pixel, but below
S/N $\sim15$ the uncertainty explodes and we can only assume that the star is
where the brightest pixel is. Open points in Figure~8 are real measurements.
They correspond to standard deviations of 100 independent centroid estimates
for 56 bright stars detected on 100 frames resampled to the same pixel grid
(before subtraction). The atmospheric limit for the accuracy of the centroid
is about 0.06 pixels in our data (Alard \& Lupton 1998). The anomalous refraction,
which depends on the color of the star, should be small in the spectral region
as red as the $I$-band, but it can only further increase this limit. Obviously,
very low error bars on the positions in the simulation are unrealistic. We note
full consistency of the results for bright stars with the scatter around the fit
to the coordinate transformation. Our conclusion from Figure~8, for variables
and stars in normal frames alike, is that for S/N$>$20 in the peak used for
centroiding the accuracy is limited by the atmosphere to 0.06--0.10 pixels
(0.024--0.040 arcsec), while below S/N $\sim15$ suddenly becomes large.
For the variables in our database, the difference light curve and the seeing data
can be used to recover the above S/N ratio, which in turn gives a rough estimate
of the centroid error.

\section{Calibration of fluxes}

\begin{figure}
\plotfiddle{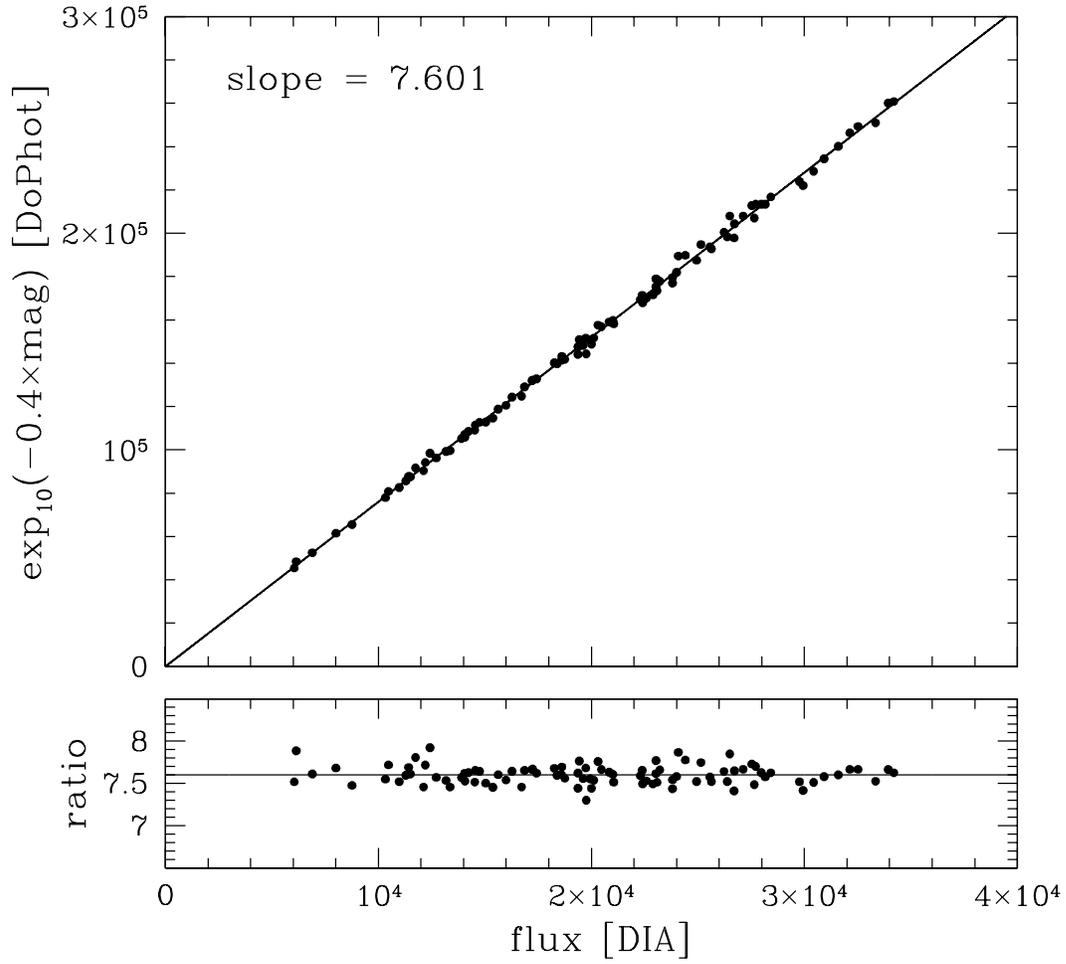}{8cm}{0.}{70.}{70.}{-233}{-100}
\caption{Transformation between raw DoPhot magnitudes and our flux units
for SC1 field. The best fit proportionality factor from 120 bright, isolated
stars is 7.601, with about 1\% uncertainty. The lower panel shows the ratio
of the data to the best fit.
}
\end{figure}

The basic type of information delivered by image subtraction is slightly
different than that from conventional photometry. Difference imaging returns
relative photometry in linear flux units, that is the light curve from which
some constant flux has been subtracted. This value depends on the brightness
of the object in question on frames used to construct the reference image.
There is no way to tell what the percentage amplitude of the variation is,
unless we obtain additional information with other means. To put the light
curve on a magnitude scale, one needs the source flux on the reference image,
the one which has been subtracted from all frames. Equation~14 helps to clarify
this concept:

\begin{equation}
m_i = C - 2.5 \log (f_0 + \Delta f_i).
\end{equation}

Difference imaging measures $\Delta f_i$, however certain applications require
$f_0$, and that must be measured separately, usually with the use of the PSF
photometry on the reference image. This has been known in the past as AC signal
vs. DC signal. Conventional PSF photometry measures $(f_0 + \Delta f_i)$
for each frame separately, however in crowded fields the noise is pushed far
above the photon statistics, and the error distribution is no longer Gaussian
due to complicated systematics of the multi-PSF fits in the presence of seeing
variations. In image subtraction, on the other hand, we can make very accurate
measurements of $\Delta f_i$, and $f_0$ will be no less accurate than
in the previous case ---  we can always run DoPhot on the reference image.

A separate issue is the zero point calibration, by which we mean the
correspondence between the computed number of counts and the flux or magnitude
in a standard photometric system. In the simplest case there will be a constant
proportionality factor between the instrumental and standard fluxes.
Image subtraction brings one little complication here. There may be a different
transformation of flux units for $\Delta f_i$ and $f_0$ due to different
normalizations of the PSF in the software or more subtle effects like differing
aperture corrections. Therefore in Equation~14 before we can set the $C$
constant, we need to ensure that $\Delta f_i$ and $f_0$ are measured consistently.

The conversion factor between our difference fluxes and the absolute DoPhot
fluxes was determined using isolated candidate variables from our catalogs,
for which we could identify a ``DoPhot'' star within 0.1 pixels. For each variable
we estimated the contaminating flux of the neighboring stars on the reference
image using coordinates of all detectable stars within 10 pixels and a Gaussian
PSF model with the approximate FWHM value. We only used variables contaminated
by less than 0.5\% according to the above estimate. For such variables the simple
photometry from our DIA catalog entry is not influenced by crowding and should
be very accurate. Only a small fraction of all stars, and even fewer candidate
variables, can satisfy this strict limit, nevertheless there are enough for
a good fit to a straight line. Figure~9 shows the data for SC1 field and
a one parameter fit to all 120 points. The best fit value of the proportionality
factor is 7.601 with 1\% uncertainty. Individual fits for all fields give about
the same scatter in the resulting slope.

The approximate zero point shift for OGLE data between raw DoPhot magnitudes
and standard $I$ band magnitudes can be useful for comparisons. The offset
of 25.6 mag added to the DoPhot output provides a crude match.

\section{Practical considerations}

\subsection{Computing power requirements}

The computer resources required for difference image photometry are not that
different from those needed for DoPhot working in fixed position mode. However,
there is plenty of room for inefficiencies if large amounts of data are
approached incorrectly. A set of 324 2k$\times$8k images (11 GB of pixel data)
takes about 5 days to process with the code compiled and {\tt -O3} optimized
using GNU {\tt gcc} compiler on Pentium III 500 MHz PC with 384 MB of RAM
memory. I do not recommend running our pipeline on comparably sized data sets
with less computing power. About 80\% of the computing time is approximately
equally shared between two procedures: the expansion of the least-squares
matrix for constant kernel to the spatially variable case, and the LU
decomposition. Profile experiments with {\tt gprof} showed that the code
is not limited by the disk or memory access.
Images taken in still frame mode would eliminate most of the
PSF variability present in the current data, and would allow much larger
subframes than ours to be processed with the same order of polynomial fits.
Because the computing time rapidly increases for higher order spatial
dependence
of the kernel, more throughput could be achieved. In the case of the main
subtracting program it is also very inefficient to process frames one by one.
The main calculation, that of the least-squares matrix, can be done once for
the entire series of images. Our pipeline can be adapted to online systems;
the reference image and the least-squares matrix would be prepared at the
beginning of the real time processing and stored. At a later time this would
allow processing a single frame as quickly as in the current setup and
facilitate nearly real time photometry.

\subsection{Availability of the data}

As an attachment to the present paper we offer the access to difference
photometry for SC1 field, the first of the OGLE-II bulge fields. The data
and the programs can be found at
{\it http://astro.princeton.edu/$\sim$wozniak/dia} . {\tt README} file explains
the details. The distribution includes a catalog of 4597 candidate variables,
the database of photometric measurements for all variables
in the catalog, a reference image (256 sections), DoPhot photometry on the
reference image, and the magnitude zero points for each of the 256 sections.
For easy access the images are FITS files (85 MB) and the photometric
measurements are {\tt gzip} compressed ASCII tables (25 MB).
All coefficients derived in this paper will apply for SC1 data
and future releases. We believe that our pipeline can still be significantly
improved, suggestions are welcome and should be sent to the author by email.
Extracting variability from difference images without allowing too many
spurious objects is a problem of particular interest. Therefore we include
with the above data a series of difference frames for 1k$\times$1k region
of the full field (1 GB), which should be a good test sample for alternative tools.

\section{Summary}

I have presented a photometric pipeline based on the Alard \& Lupton
optimal image subtraction algorithm and capable of processing very large
data sets in automated fashion, as is necessary in microlensing searches.
The performance was evaluated based on complete reanalysis of 3 observing
seasons of data for 49 Galactic bulge fields, amounting to 11,000 images
or 380 GB. The photometry obtained in the course of this project will be
gradually published. Currently the data for one of the bulge fields is
available from anonymous ftp, and we plan to release the data for the remaining
fields. Perhaps for the first time, a very weakly filtered large sample of suspected
variable stars is publicly available for detailed studies.
The overall noise properties are exceptionally good for this
kind of massive variability search in crowded fields. The error distribution
is nearly Gaussian with $\sim0.7$\% of the points in the non-Gaussian wings.
The scatter in the photometry for the faint stars is close to 1.17 times photon
noise. For brighter stars this ratio rises due to systematics, nevertheless
compared with DoPhot, our photometry is at least 2--3 times better over
the entire range of the observed magnitudes. At the bright end, $I\/\sim$11--13 mag,
the random errors are about 0.005 mag. Sensitivity to microlensing and
the amount of information that can be inferred from statistical analyses
improve accordingly. The most striking example is a factor of 2 increase in the
number of detected microlensing events. The modular structure of the pipeline
provides a sound basis for future development of the implementation.

\acknowledgments

I thank Prof. Bohdan Paczy\'nski for encouragement and support in this work,
and Christophe Alard \& Robert Lupton for numerous discussions on image
processing. I used many programming solutions invented by Christophe Alard.
Also I would like to thank Prof. Andrzej Udalski and the OGLE team
for the full access to the OGLE-II databases prior to publication.
Igor Soszy\'nski kindly provided an improved version of program {\tt resample}.
This work was supported with NSF grant AST--9820314 to Bohdan Paczy\'nski.

\end{document}